\documentclass[11pt]{article}

\usepackage[utf8]{inputenc}
\usepackage[T1]{fontenc}
\usepackage{lmodern}
\usepackage[margin=1in]{geometry}
\usepackage{booktabs}
\usepackage{amsmath}
\usepackage{array}
\usepackage{enumitem}
\usepackage[hidelinks]{hyperref}
\usepackage{url}

\setlist{itemsep=2pt, topsep=4pt}
\newcommand{\cfg}[1]{\textsc{#1}}
\newcommand{\sparks}{\texttt{spark-s1}}

\title{\textbf{ Beyond Plain MCP: Do Decision Models and Code Mode Make Tool Use Cheaper and Faster?}\\
A Pre-Registered Study of Code Mode and Decision Layers in MCP Gateways}

\author{Abhishek Rai\\
Independent researcher\\
\small Models and data: \url{https://huggingface.co/abhishek085}}
\date{November 2026}

\begin{document}
\maketitle

\begin{abstract}
Model Context Protocol (MCP) gateways sit between an agent and a growing catalog of tools. Two designs are popular for keeping that catalog from swamping the agent: \emph{code mode}, where the agent writes a script against tool stubs and only the final output re-enters its context, and a \emph{decision layer}, a small model that picks the tools before the agent sees them. We ask when each pays off and whether they combine.
We build a controlled harness (a 12-server, 42-tool mock MCP world plus up to 1{,}958 real distractor tools; four gateway configurations; three open-weight agents: Qwen3.8-27B, Gemma-4-26B-A4B, Nemotron-3-Nano-30B-A3B) and report a \emph{pre-registered evaluation on a sealed test split} (119 instances, 1{,}785 serial runs, Holm correction over 12 comparisons), a pre-specified \emph{extension} (119 further unseen instances, a stub ablation, a bootstrap that resamples task families), and exploratory development-set analyses.
(i)~Code mode beats plain MCP on utility for Qwen ($\Delta U=+0.364$) and Nemotron ($+0.861$), robustly to clustering, with 70--90\% fewer tokens. (ii)~Telling the agent what each tool \emph{returns} (return-shape hints in the stubs) matters a great deal for two of the three agents: removing the hints lowers code-mode success by $22.7$ points (Gemma) and $39.5$ points (Nemotron), Holm-adjusted $p=0.0003$, but not significantly for Qwen ($1.7$ points). Those effects on utility ($+0.30$, $+0.48$) are larger than the effect of adding a router. (iii)~The central hypothesis, that a decision layer in front of code mode beats code mode alone, is \emph{not confirmed}: pooled over 238 instances the effects are $-0.028$ (Qwen), $+0.099$ (Gemma) and $+0.109$ (Nemotron), none significant after correction, although the signs agree across two independent samples. (iv)~Adding code mode to a router beats the router alone for Qwen under family-level resampling. Exploratory analyses add that shape routing never beat ``always code with a fallback'' and that an unbounded fallback made the decision layer $14\times$ slower at 2{,}000 tools. We release the harness, a fine-tuned 4B tool-selection model and all run records.
\end{abstract}

\section{Introduction}

An agent that talks to hundreds of MCP tools pays for them twice. Every tool schema is placed in the prompt on every turn, and every tool result is copied back through the model on its way to the next call. Two remedies have emerged. \emph{Code mode} replaces JSON tool calls by a script that the agent writes against generated stubs; the script runs in a sandbox, chains calls, filters large results, and returns one short answer \cite{cloudflare-codemode,anthropic-codeexec}. A \emph{decision layer} puts a cheap model in front of the agent to choose which tools to expose, in the spirit of tool retrieval and routing \cite{ragmcp,mcpzero,aceRouter}.

Recent controlled work shows that code-style tool calling matches or beats JSON calling for most models \cite{bitterlesson,executecode,felendler}, and tool retrieval is established \cite{ragmcp,mcpzero}. What is less clear is how the two interact, and which of the many design details (the stubs, the router, the fallback) carries the effect. We started from the natural hypothesis that a decision layer in front of code mode would beat either alone, pre-registered it, and tested it on a sealed split. The data told a more nuanced story, and this report documents it, including the negative results.

\paragraph{Contributions.}
\begin{enumerate}
\item A controlled four-configuration comparison (plain MCP, router+JSON, code mode, router+code mode) across three open-weight agents, with a \emph{pre-registered confirmatory evaluation} on a sealed split (matched cells, paired bootstrap, Holm correction, serial latency; Section~\ref{sec:confirm}), and exploratory development-set analyses (Section~\ref{sec:results}).
\item Evidence that the \emph{return-shape information in code-mode stubs} is a first-order variable: on the sealed split, removing it lowers success by 22.7 and 39.5 points for two of three agents (Section~\ref{sec:ext1}), replicating the development-set result (Section~\ref{sec:stubs}). To our knowledge, no prior controlled study isolates this variable (see Section~\ref{sec:related}).
\item A characterization of \emph{when} a decision layer helps: its benefit in front of code mode is directionally positive for the two lower-performing agents, absent for the strongest, and not statistically confirmed even with 238 instances (Sections~\ref{sec:confirm}, \ref{sec:decision}); it never needs to choose the call shape, and it can be dramatically worse without a bounded fallback (Section~\ref{sec:scale}).
\item A 4B tool-selection model fine-tuned from the \sparks{} decision model, with a measured recall, latency and robustness profile, including the finding that a prompted-LLM router is fully steerable by a poisoned tool description (Sections~\ref{sec:router}--\ref{sec:security}).
\item A frank account of what is \emph{not} established (Section~\ref{sec:limits}).
\end{enumerate}

\section{Related Work}\label{sec:related}

\paragraph{Tool use and code as action.}
Tool-augmented language models range from self-supervised API calls \cite{toolformer} and interleaved reasoning and acting \cite{react} to models trained on thousands of real APIs \cite{toolllm,gorilla}. Letting the agent emit executable code rather than structured calls was shown to improve agents in CodeAct \cite{codeact}, and LLMCompiler plans and executes function calls as a dependency graph to parallelize them \cite{llmcompiler}. MCP itself standardizes how tools are described and invoked \cite{mcp}; benchmarks such as MCP-Bench \cite{mcpbench} and MCPToolBench++ \cite{mcptoolbenchpp} evaluate agents against MCP servers.

\paragraph{Code mode for MCP.}
Cloudflare's ``Code Mode'' and Anthropic's code execution with MCP and programmatic tool calling expose MCP tools as a code API so that intermediate data stays out of the model context \cite{cloudflare-codemode,anthropic-codeexec,anthropic-advanced}. Controlled comparisons now exist: Patel et al.~\cite{bitterlesson} compare code-based and JSON tool calling for 14 models on BFCL v4 and find code matches or beats JSON for 11 of 14; Yang et al.~\cite{executecode} ablate restricting a coding agent to an \texttt{execute\_code} tool and report cost gains with tied pass rates; Felendler et al.~\cite{felendler} study MCP design choices on MCP-Bench and report token and latency savings together with a larger attack surface. These works establish that code mode works and that it varies by model. We ask \emph{why it varies}, and show that in our setting a single stub-design variable (declared return shapes) changes code-mode success by up to 40 points for some agents and not for others.

\paragraph{Tool retrieval, routing and discovery.}
RAG-MCP retrieves relevant MCP tools by semantic search to avoid prompt bloat \cite{ragmcp}; MCP-Zero lets the agent actively request tools in a hierarchical server-then-tool process \cite{mcpzero}; ACE-Router trains history-aware routers for large MCP and agent ecosystems \cite{aceRouter}; and Anthropic's tool search defers loading of tool definitions \cite{anthropic-advanced}. Our router configurations (\cfg{B}, \cfg{D}) are in this family; we differ in (a) pairing the router with code mode, (b) measuring end-to-end latency and fallback behavior rather than only recall, and (c) showing that recall is capped by the candidate retriever at large catalogs. More general model routing \cite{routellm} and per-step routing for agent systems \cite{evoroute} choose among \emph{models}; our decision layer chooses tools and call shape.

\paragraph{Speculation, caching and skills.}
Several systems reduce agent latency by overlapping or speculating on tool execution \cite{specactions,paste}, or by reusing learned tool compositions \cite{skillcraft}. These are orthogonal to our question and could be layered on a code-mode gateway; we do not evaluate them.

\paragraph{Security of tool selection and tool descriptions.}
ToolHijacker shows that an injected tool document can hijack tool selection in agents \cite{toolhijacker}; MCP-ITP automates implicit tool poisoning in MCP \cite{mcpitp}; TRUSTDESC generates trusted tool descriptions as a defense \cite{trustdesc}. Our router-attack result (Section~\ref{sec:security}) is a small extension of this line: it shows that the \emph{router} is itself the target, and that a scoring router behaves differently from a prompted one.

\paragraph{Failure recovery and static checking.}
Runtime failure detection and repair for agents \cite{rtrepair} and budgeted recovery policies in tool-augmented orchestrators \cite{selfheal} motivate the fallback analysis in Section~\ref{sec:scale}. Static analysis of generated code against known APIs can catch hallucinated library calls \cite{staticanalysis}; typed stubs in code mode make a similar check possible for tool scripts, which we identify as future work.

\section{Study Design}\label{sec:setup}

\subsection{Gateway configurations}
A single gateway process mediates between the agent and the MCP servers; the four configurations are switches on it, so servers, tasks, seeds, turn caps and sandbox are held constant.

\begin{table}[h]
\centering\small
\begin{tabular}{@{}llll@{}}
\toprule
Id & Tool choice & Call shape & Seen by the agent\\
\midrule
\cfg{A} Plain MCP & agent, from all schemas & JSON calls & all schemas; every result\\
\cfg{B} Router + JSON & decision model (subset) & JSON calls & subset schemas; every result\\
\cfg{C} Code mode & agent via \texttt{search()} & script in sandbox & 2 meta-tools; final line\\
\cfg{D} Router + code & decision model (subset) & script; JSON fallback & subset stubs; final line\\
\bottomrule
\end{tabular}
\caption{The four gateway configurations. \cfg{D-oracle} uses the gold tool list at zero router cost.}
\end{table}

\paragraph{Code mode.} The agent has two meta-tools: \texttt{search()} returns Python stubs for matching tools, and \texttt{execute()} runs a script in a sandbox (an rlimit-restricted subprocess, not a container) with up to three executions per task. \emph{Return-shape hints} are hand-written descriptions of each tool's return value (keys, types, list versus dict) appended to the stub; in the \emph{unhinted} setting a stub only states \texttt{-> dict | list}.
\paragraph{Decision layer.} Given the task and a candidate list, the router returns a tool subset. Variants: a prompted LLM router (Gemma for the Qwen agent, Qwen for the Gemma agent, never the agent itself); zero-shot \sparks{}; and our fine-tuned \sparks{}-based router (Section~\ref{sec:router}). \cfg{D} may also choose the call shape by rules or a learned classifier. On a subset miss, the default policy retries in code mode with search over the \emph{full} catalog and, only if that fails, falls back to JSON over the subset with \texttt{request\_more\_tools} capped at two calls (\texttt{full\_code} policy). An earlier unbounded policy (\texttt{json\_more}) is reported as a baseline.

\subsection{World, tasks and agents}
The mock world has 12 servers and 42 tools (issues, chat, billing, logs, CRM, docs, files, calendar, deployments, inventory, HR, support) with seeded, resettable state; scale arms add up to 1{,}958 real distractor tools drawn from a public catalog, for catalogs of 42, 200 and 2{,}000 tools. Tasks come from programmatic template families (lookup chains, aggregation over logs and invoices, conditional actions with side effects, multi-server workflows). Each kept task is solvable by a reference solver and verified by inspecting world state, not model text. The development split is seeds 0--2 with salt \texttt{dev} (119 task instances).

Agents are Qwen3.8-27B (checkpoint \texttt{Vontra/Qwen3.8-27B-MLX-4bit}, 4-bit, served with the TensorFold runtime), Gemma-4-26B-A4B (\texttt{nvidia/Gemma-4-26B-A4B-NVFP4}, vLLM) and Nemotron-3-Nano-30B-A3B (\texttt{nvidia/NVIDIA-Nemotron-3-Nano-30B-A3B-NVFP4}, vLLM), all quantized community or vendor checkpoints, served locally on one NVIDIA GB10 system (121\,GB unified memory); the router runs in bf16 under vLLM. Temperature is 0 and thinking is disabled for all agents; one agent model generates at a time.

\subsection{Metrics and statistics}
Success is the verifier's pass. Tokens are total prompt and completion tokens including router tokens. Latency is measured only in serial passes (concurrency 1, nothing else generating). Utility is
\[
U = \text{success} - \lambda_t \cdot \tfrac{\text{tokens}}{1000} - \lambda_\ell \cdot \text{latency}_s ,
\]
with $\lambda_t=0.01$ and $\lambda_\ell=0.002$ as primary weights (chosen before seeing \cfg{D} and \cfg{B} results; $\lambda_\ell=0.02$ is also reported). Where latency was not measured serially, we report $U_{\text{tok}}$ (token term only). All intervals are 95\% paired bootstrap intervals over matched task-seed cells.

\subsection{Pre-registration and what is confirmatory}
A written protocol is in the repository (\texttt{docs/PREREG.md}). Its first part was written \emph{after} the first pilots and is flagged as such; everything produced under it (development seeds, harness changes during development) is \textbf{exploratory}. A confirmatory protocol was committed before any run on the sealed test split and the configuration was frozen at git tag \texttt{confirmatory-freeze-v1} (the pre-registration commit; the harness was not modified afterwards, only a serial driver script and the analysis scripts were added). The protocol fixes: salt \texttt{test}, seeds 0--2, 119 task instances (41/38/40 after a quality filter; every kept task is solvable by its reference solver and deterministic, which we re-verified without reading any agent outcome), all three agents, configurations \cfg{A}, \cfg{B}, \cfg{C}, \cfg{D} and \cfg{D-oracle}, serial execution, utility $U$ with $\lambda_t=0.01$ and $\lambda_\ell=0.002$ as the primary endpoint, paired bootstrap (10{,}000 resamples), Holm correction over the 12 primary comparisons (H1--H4 per agent), and the rule that a run exceeding the 900\,s cap counts as a failure.

\paragraph{Sample size.} The protocol itself states that about 300 task-seed pairs per arm are needed to resolve the 2-point non-inferiority margin. The confirmatory split has 119 instances, so that margin cannot be resolved and ``not confirmed'' below means \emph{inconclusive}, not absent.

\paragraph{Extension protocol.} After the core confirmatory results had been analysed, and before any further run, we wrote an extension protocol in the same file. It specifies: (Ext-1) code mode \emph{without} return-shape hints on the same 119 test instances, to test the stub hypothesis on the sealed split; (Ext-2) configurations \cfg{C}, \cfg{D} and \cfg{D-oracle} on 119 \emph{new} test instances (seeds 3--5; solvable and deterministic, checked beforehand), to test H3 on 238 instances in total, still below 300; and (Ext-3) a bootstrap that resamples task \emph{families} rather than instances, plus a sensitivity grid over the utility weights. These are labelled extensions and are outside the 12-comparison Holm family (Ext-1 and Ext-2 each carry their own Holm correction over three agents). One deviation: before the core run we made a single smoke run on one test-split task to check the services (not part of the results).

\section{Confirmatory Results (Sealed Test Split)}\label{sec:confirm}

\paragraph{Setup.} All numbers in this section come from the frozen protocol: 3 agents $\times$ 5 configurations $\times$ 119 instances $=1{,}785$ serial runs (concurrency 1, one agent model generating at a time), hinted stubs, thinking disabled, temperature 0. In contrast to the development runs, \cfg{B} and \cfg{D} use the fine-tuned \sparks{} router (Section~\ref{sec:router}; top-30 embedding candidates, threshold 0.05) for \emph{all} agents. \cfg{D-oracle} receives the gold tool list at zero router cost and serves as a reference, not a primary test. Table~\ref{tab:conf} gives the per-configuration results and Table~\ref{tab:holm} the 12 pre-registered comparisons. Across the 1{,}785 runs there were no hangs and one infrastructure fault (a malformed router response on one Gemma \cfg{D} run), which was re-run once under the identical configuration as the protocol prescribes; the original record is retained in the release.

\begin{table}[h]
\centering\small
\begin{tabular}{@{}llrrrr@{}}
\toprule
Agent & Config & Success & Tokens/task & Median latency (s) & $U$\\
\midrule
Qwen3.8-27B & \cfg{A} plain MCP & 0.857 & 26{,}602 & 4.9 & 0.576\\
 & \cfg{B} router+JSON & 0.824 & 15{,}099 & 5.0 & 0.648\\
 & \cfg{C} code mode & \textbf{0.992} & 4{,}362 & 3.6 & \textbf{0.940}\\
 & \cfg{D} router+code & 0.983 & 5{,}703 & 4.7 & 0.916\\
 & \cfg{D-oracle} (reference) & 1.000 & 3{,}083 & 3.1 & 0.962\\
\midrule
Gemma-4-26B-A4B & \cfg{A} & 0.866 & 16{,}779 & 5.7 & 0.680\\
 & \cfg{B} & 0.866 & 10{,}439 & 7.0 & 0.741\\
 & \cfg{C} & 0.891 & 5{,}112 & 6.4 & 0.798\\
 & \cfg{D} & \textbf{0.933} & \textbf{3{,}921} & 8.1 & \textbf{0.875}\\
 & \cfg{D-oracle} (reference) & 1.000 & 2{,}098 & 6.1 & 0.963\\
\midrule
Nemotron-3-Nano & \cfg{A} & 0.361 & 74{,}608 & 6.2 & $-0.404$\\
 & \cfg{B} & 0.563 & 35{,}547 & 6.1 & 0.167\\
 & \cfg{C} & 0.555 & 7{,}553 & 4.6 & 0.457\\
 & \cfg{D} & \textbf{0.672} & 9{,}867 & 5.3 & \textbf{0.552}\\
 & \cfg{D-oracle} (reference) & 0.681 & 9{,}037 & 4.0 & 0.571\\
\bottomrule
\end{tabular}
\caption{Confirmatory results, sealed test split, 119 matched instances per cell. $U=\text{success}-0.01\,\text{tokens}/1000-0.002\,\text{latency}_s$ (mean). Medians are shown because a few runs are very long (Gemma \cfg{C}: one 890\,s run and three others above 60\,s, so its mean latency is 20.9\,s versus a 6.4\,s median).}
\label{tab:conf}
\end{table}

\begin{table}[h]
\centering\scriptsize
\begin{tabular}{@{}llrrrrr@{}}
\toprule
 & & & \multicolumn{2}{c}{Pre-registered (instances)} & \multicolumn{2}{c}{Resampling families}\\
\cmidrule(lr){4-5}\cmidrule(lr){6-7}
Agent & Comparison & $\Delta U$ & 95\% CI & Holm $p$ & 95\% CI & Holm $p$\\
\midrule
Qwen & H1: \cfg{C} $-$ \cfg{A} & $+0.364$ & $[+0.251,+0.504]$ & 0.0012 & $[+0.158,+0.654]$ & 0.0012\\
Qwen & H2: \cfg{B} $-$ \cfg{A} & $+0.072$ & $[-0.111,+0.230]$ & 0.43 & $[-0.116,+0.223]$ & 1.00\\
Qwen & H3: \cfg{D} $-$ \cfg{C} & $-0.025$ & $[-0.056,+0.005]$ & 0.43 & $[-0.073,+0.010]$ & 1.00\\
Qwen & H4: \cfg{D} $-$ \cfg{B} & $+0.268$ & $[+0.133,+0.451]$ & 0.0012 & $[+0.070,+0.502]$ & 0.042\\
Gemma & H1 & $+0.118$ & $[-0.011,+0.244]$ & 0.43 & $[-0.151,+0.357]$ & 1.00\\
Gemma & H2 & $+0.061$ & $[-0.028,+0.143]$ & 0.43 & $[-0.048,+0.172]$ & 1.00\\
Gemma & H3 & $+0.077$ & $[-0.009,+0.176]$ & 0.43 & $[-0.037,+0.272]$ & 1.00\\
Gemma & H4 & $+0.133$ & $[+0.050,+0.222]$ & 0.0084 & $[-0.001,+0.309]$ & 0.43\\
Nemotron & H1 & $+0.861$ & $[+0.674,+1.046]$ & 0.0012 & $[+0.470,+1.184]$ & 0.0012\\
Nemotron & H2 & $+0.571$ & $[+0.378,+0.754]$ & 0.0012 & $[+0.208,+0.904]$ & 0.045\\
Nemotron & H3 & $+0.095$ & $[-0.008,+0.199]$ & 0.43 & $[-0.077,+0.268]$ & 1.00\\
Nemotron & H4 & $+0.385$ & $[+0.175,+0.616]$ & 0.0048 & $[-0.067,+0.842]$ & 0.69\\
\bottomrule
\end{tabular}
\caption{The 12 pre-registered primary comparisons (endpoint $U$, $\lambda_\ell=0.002$, 10{,}000 bootstrap resamples, Holm over 12). Left: the pre-registered analysis, which resamples instances. Right: an extension that resamples the 21 task families, which respects the clustering of instances within families.}
\label{tab:holm}
\end{table}

\paragraph{Instances are clustered.} The 119 instances are instantiations of only 21 task templates, so they are not independent. The pre-registered bootstrap resamples instances; we therefore also resample families (right-hand columns of Table~\ref{tab:holm}). The family-level intervals are wider, and two of the six instance-level significant results (H4 for Gemma and for Nemotron) do not survive Holm correction. We treat the family-level analysis as the more defensible one and phrase conclusions accordingly, while keeping the pre-registered analysis in the table.

\subsection{Primary comparisons}\label{sec:primary}
\paragraph{H1, code mode versus plain MCP: supported for Qwen and Nemotron under both analyses.} Code mode cuts tokens by 84\% (Qwen), 70\% (Gemma) and 90\% (Nemotron). Success rises by $+13.4$ points $[+7.6,+20.2]$ for Qwen and $+19.3$ points $[+8.4,+30.3]$ for Nemotron. For Gemma the direction is the same ($+2.5$ points of success) but the interval includes zero under both analyses.

\paragraph{H2, router+JSON versus plain MCP: supported only for Nemotron} (Holm $p=0.0012$ and $0.045$ under the two analyses; $+20.2$ points of success). For Qwen and Gemma the router cut tokens but not utility significantly. The 2-point non-inferiority margin is unresolvable at this $n$.

\paragraph{H3, router+code versus code mode alone: not confirmed.} This is the central hypothesis. Point estimates are $-0.025$ (Qwen), $+0.077$ (Gemma) and $+0.095$ (Nemotron); none is significant (raw $p=0.07$--$0.10$, Holm $p=0.43$). Secondary success differences are $-0.8$ points (Qwen), $+4.2$ points $[-1.7,+10.1]$ (Gemma) and $+11.8$ points $[+3.4,+21.0]$ (Nemotron); the last is unadjusted and hypothesis-generating. Section~\ref{sec:ext2} adds 119 unseen instances.

\paragraph{H4, router+code versus router+JSON: supported for Qwen under both analyses.} At the instance level it holds for all three agents (Holm $p\le0.0084$); at the family level only Qwen survives ($p=0.042$), Gemma is borderline (CI $[-0.001,+0.309]$) and Nemotron's interval is wide. The direction is positive for all three: once a router has produced a small subset, executing scripts is better than JSON calls, with large token savings.

\paragraph{Headroom and router quality.} \cfg{D-oracle} receives the reference solver's tool list, which leaks the task plan, so it is a ceiling, not a realistic router. It improves on \cfg{C} by $+0.022$ $[+0.009,+0.042]$ for Qwen, $+0.165$ $[+0.083,+0.263]$ for Gemma and $+0.114$ $[+0.001,+0.227]$ for Nemotron. The fine-tuned router's complete recall on the test split is 0.975 (3 of 119 instances miss a needed tool; mean recall 0.987) at 2.6 exposed tools and 1.55--1.61\,s per decision; the JSON fallback fired in 1.7\%, 0\% and 13.4\% of \cfg{D} runs. We did not decompose the remaining gap between \cfg{D} and \cfg{D-oracle}.

\subsection{Extension 1: do return-shape hints matter on the sealed split?}\label{sec:ext1}
Ext-1 reruns code mode on the same 119 test instances with the hints removed (stubs state only \texttt{-> dict | list}) and compares it with the hinted \cfg{C} of Table~\ref{tab:conf}. Intervals resample families; the Holm correction covers the three agents.

\begin{table}[h]
\centering\small
\begin{tabular}{@{}lrrrrr@{}}
\toprule
Agent & Success, no hints & Success, hints & $\Delta$ success [95\% CI] & $\Delta U$ [95\% CI] & Holm $p$\\
\midrule
Qwen3.8-27B & 0.975 & 0.992 & $+0.017$ $[-0.023,+0.080]$ & $+0.035$ $[-0.010,+0.107]$ & 0.20\\
Gemma-4-26B-A4B & 0.664 & 0.891 & $+0.227$ $[+0.119,+0.361]$ & $+0.296$ $[+0.162,+0.459]$ & 0.0003\\
Nemotron-3-Nano & 0.160 & 0.555 & $+0.395$ $[+0.190,+0.571]$ & $+0.478$ $[+0.235,+0.693]$ & 0.0003\\
\bottomrule
\end{tabular}
\caption{Extension 1: effect of return-shape hints on code-mode success and utility, sealed test split (119 instances, 21 families). The hints also cut tokens: $-1{,}598$ (Qwen), $-4{,}372$ (Gemma), $-6{,}104$ (Nemotron), all intervals excluding zero.}
\label{tab:ext1}
\end{table}

The development-set result (Section~\ref{sec:stubs}) replicates almost exactly: without hints Gemma and Nemotron fall to success 0.66 and 0.16 (development: 0.66 and 0.19), well below plain MCP, while Qwen is barely affected. On a common scale ($\Delta U$), return shapes are worth $+0.30$ and $+0.48$ for the two lower-performing agents, whereas adding a router in front of code mode is worth $+0.10$ and $+0.11$ and is not significant (Section~\ref{sec:ext2}). For Qwen neither matters much. The hints are hand-written by the author, so this shows that \emph{information about return values} matters, not that any particular way of producing it works.

\subsection{Extension 2: H3 with more instances}\label{sec:ext2}
Ext-2 runs \cfg{C}, \cfg{D} and \cfg{D-oracle} on 119 new test instances (seeds 3--5), with the frozen configuration, for all three agents. We report the new sample alone and pooled with the original (238 instances); Holm correction covers the three pooled tests.

\begin{table}[h]
\centering\small
\begin{tabular}{@{}llrrrrr@{}}
\toprule
Agent & Sample & $n$ & \cfg{C} success & \cfg{D} success & $\Delta U$ (\cfg{D}$-$\cfg{C}) [95\% CI] & Holm $p$\\
\midrule
Qwen & seeds 3--5 & 119 & 1.000 & 0.983 & $-0.032$ $[-0.076,-0.007]$ & \\
 & pooled & 238 & 0.996 & 0.983 & $-0.028$ $[-0.075,+0.000]$ & 0.15\\
Gemma & seeds 3--5 & 119 & 0.866 & 0.958 & $+0.121$ $[+0.010,+0.281]$ & \\
 & pooled & 238 & 0.878 & 0.945 & $+0.099$ $[-0.011,+0.273]$ & 0.21\\
Nemotron & seeds 3--5 & 119 & 0.504 & 0.681 & $+0.123$ $[-0.091,+0.343]$ & \\
 & pooled & 238 & 0.529 & 0.676 & $+0.109$ $[-0.071,+0.294]$ & 0.22\\
\bottomrule
\end{tabular}
\caption{Extension 2: H3 on 119 unseen test instances and pooled with the original 119. Intervals resample the 21 task families. The ``seeds 3--5'' rows are unadjusted descriptive replications.}
\label{tab:ext2}
\end{table}

H3 remains \emph{not confirmed}: after pooling, no agent is significant after correction. The new sample reproduces the signs seen in the original sample: negative for Qwen (the router costs about 1\,s and a little success), positive for Gemma and Nemotron. For Gemma the new sample alone has an interval above zero ($+9.2$ points of success, $+0.121$ in $U$) before adjustment; for Nemotron the effect is similar in size but uncertain. We read this as moderate evidence for a benefit of a decision layer for agents whose own code mode is weaker, and for a small cost for the strongest agent, not as confirmation.

\subsection{Latency and scale}\label{sec:conflatency}
For Qwen, \cfg{D} is $1.4$\,s slower than \cfg{C} (mean difference $+1.45$\,s $[+1.16,+1.79]$) while the oracle router makes it $0.4$\,s faster, reproducing the development finding that a front-door router must cost less than about 0.4--0.7\,s to pay for itself on a fast agent. For Gemma, differences in mean latency are dominated by a few runaway \cfg{C} runs; the typical latency of \cfg{D} is higher (median 8.1\,s versus 6.4\,s) but its tail is shorter (p90 15.0\,s versus 25.4\,s). With $\lambda_\ell=0.02$ the Gemma \cfg{D}$-$\cfg{C} difference is $+0.282$ $[+0.018,+0.699]$, driven by the tail.

\begin{table}[h]
\centering\small
\begin{tabular}{@{}rlrrr@{}}
\toprule
Tools & Config & Success & Tokens/task & Latency mean / median / p90 (s)\\
\midrule
42 & \cfg{A} & 0.90 & 16{,}076 & 3.4 / 3.3 / 5.5\\
42 & \cfg{C} & 1.00 & 4{,}112 & 3.3 / 3.0 / 4.9\\
42 & \cfg{D} & 1.00 & 5{,}045 & 4.6 / 4.5 / 5.9\\
200 & \cfg{A} & 0.80 & 148{,}047 & 20.5 / 30.2 / 33.1\\
200 & \cfg{C} & 1.00 & 7{,}766 & 4.2 / 4.4 / 6.6\\
200 & \cfg{D} & 1.00 & 15{,}176 & 9.3 / 5.5 / 19.1\\
2{,}000 & \cfg{A} & 0.00 (context overflow) & -- & --\\
2{,}000 & \cfg{C} & 0.90 & 11{,}942 & 6.0 / 5.0 / 14.7\\
2{,}000 & \cfg{D} & 0.90 & 31{,}016 & 16.8 / 16.6 / 38.0\\
\bottomrule
\end{tabular}
\caption{Confirmatory scale arm (Qwen3.8-27B, test seed 0, 10 tasks, serial; \cfg{D} uses the bounded fallback). Descriptive only.}
\label{tab:scaleconf}
\end{table}

Table~\ref{tab:scaleconf} is descriptive (10 tasks). It reproduces the development picture: code mode stays nearly flat in tokens and latency from 42 to 2{,}000 tools; plain MCP needs 148K tokens per task at 200 tools and fails by context overflow at 2{,}000; and \cfg{D} with the bounded fallback matches \cfg{C}'s success but is slower (9.3 versus 4.2\,s at 200 tools, 16.8 versus 6.0\,s at 2{,}000) and uses more tokens, because retrieval-limited recall forces full-catalog retries.

\subsection{Sensitivity to the utility weights}\label{sec:lambda}
The weights in $U$ are a choice. Table~\ref{tab:lambda} shows $\Delta U$ for H1 and H3 under five settings.

\begin{table}[h]
\centering\small
\begin{tabular}{@{}llrrrrr@{}}
\toprule
 & & $\lambda_t{=}.005$ & $\lambda_t{=}.01$ & $\lambda_t{=}.01$ & $\lambda_t{=}.01$ & $\lambda_t{=}.02$\\
Agent & Comparison & $\lambda_\ell{=}.002$ & $\lambda_\ell{=}0$ & $\lambda_\ell{=}.002$ & $\lambda_\ell{=}.02$ & $\lambda_\ell{=}.002$\\
\midrule
Qwen & H1 \cfg{C}$-$\cfg{A} & $+0.253^*$ & $+0.357^*$ & $+0.364^*$ & $+0.433^*$ & $+0.587^*$\\
Qwen & H3 \cfg{D}$-$\cfg{C} & $-0.018$ & $-0.022$ & $-0.025$ & $-0.051^*$ & $-0.038$\\
Gemma & H1 & $+0.059$ & $+0.142$ & $+0.118$ & $-0.099$ & $+0.234$\\
Gemma & H3 & $+0.071$ & $+0.054$ & $+0.077$ & $+0.282$ & $+0.089$\\
Nemotron & H1 & $+0.526^*$ & $+0.864^*$ & $+0.861^*$ & $+0.835^*$ & $+1.532^*$\\
Nemotron & H3 & $+0.106$ & $+0.095$ & $+0.095$ & $+0.096$ & $+0.072$\\
\bottomrule
\end{tabular}
\caption{$\Delta U$ on the original 119 instances under different weights; $^*$: family-resampled 95\% CI excludes zero (unadjusted). The primary setting is the middle column.}
\label{tab:lambda}
\end{table}

H1 for Qwen and Nemotron holds under every setting; Gemma's H1 changes sign at $\lambda_t=0.01$, $\lambda_\ell=0.02$, because of its slow code-mode runs. H3 is positive for Gemma and Nemotron and negative for Qwen under all five settings, and is never family-significant except for Qwen at a high latency weight, where it is negative. The qualitative conclusions do not depend on the weights.

\section{Exploratory Results on Development Seeds}\label{sec:results}
The analyses in this section use development seeds (salt \texttt{dev}), one run per cell, prompted LLM routers in \cfg{B} and \cfg{D} unless noted, and a harness that changed during development. They generated the hypotheses tested in Section~\ref{sec:confirm}; they are exploratory.

\subsection{Code mode is the dominant effect}\label{sec:h1}
Table~\ref{tab:main} gives the development results with hinted stubs (matched cells: 119 for Qwen, 118 for Gemma, 119 for Nemotron).

\begin{table}[h]
\centering\small
\begin{tabular}{@{}llrrr@{}}
\toprule
Agent & Config & Success & Tokens/task & $U_{\text{tok}}$\\
\midrule
Qwen3.8-27B & \cfg{A} plain MCP & 0.832 & 30{,}649 & 0.525\\
 & \cfg{B} router+JSON (LLM router) & 0.857 & 10{,}815 & 0.749\\
 & \cfg{C} code mode & \textbf{0.975} & 4{,}474 & \textbf{0.930}\\
 & \cfg{D} router+code (LLM router) & 0.966 & 4{,}576 & 0.921\\
\midrule
Gemma-4-26B-A4B & \cfg{A} & 0.831 & 15{,}282 & 0.678\\
 & \cfg{B} (Qwen router) & 0.839 & 7{,}886 & 0.760\\
 & \cfg{C} & 0.864 & 4{,}255 & 0.822\\
 & \cfg{D} (Qwen router) & \textbf{0.966} & \textbf{3{,}487} & \textbf{0.931}\\
\midrule
Nemotron-3-Nano & \cfg{A} & 0.412 & 80{,}936 & --\\
 & \cfg{C} & 0.504 & 8{,}187 & --\\
\bottomrule
\end{tabular}
\caption{Development seeds, hinted stubs, one run per cell. \cfg{B}/\cfg{D} use a prompted LLM router (different family from the agent) with router tokens counted. Nemotron's decision-layer configurations were run only on the 39-task latency set (Table~\ref{tab:latency}).}
\label{tab:main}
\end{table}

\textbf{H1 (code mode beats plain MCP) is supported for all three agents on development seeds (confirmed for two of three on the test split)}, with a size that depends on the agent. For Qwen, \cfg{C} beats \cfg{A} by $+14.3$ points of success $[+7.6,+21.8]$ and $\Delta U_{\text{tok}}=+0.405$ $[+0.252,+0.607]$ at one-seventh of the tokens. For Gemma the success gain is small ($+3.4$ points) but the token saving is $3.6\times$ ($\Delta U_{\text{tok}}=+0.144$ $[+0.047,+0.246]$). For Nemotron, success rises from 0.41 to 0.50 with $10\times$ fewer tokens; plain MCP failed several long log-paging tasks that exceed the context window.

\subsection{The stubs matter as much as the architecture}\label{sec:stubs}
Our first runs gave code-mode stubs no return information. Nemotron's code mode then looked very weak (success 0.185 versus 0.412 for plain MCP). Adding hand-written return-shape hints to the stubs, with no other change, produced the following.

\begin{table}[h]
\centering\small
\begin{tabular}{@{}lrrrr@{}}
\toprule
Agent & \cfg{C} unhinted & \cfg{C} hinted & $\Delta$ success & \cfg{A} (reference)\\
\midrule
Qwen3.8-27B & 0.950 & 0.975 & $+2.5$ pts & 0.832\\
Gemma-4-26B-A4B & 0.664 & 0.864 & $+20.0$ pts & 0.831\\
Nemotron-3-Nano & 0.185 & 0.504 & $+31.9$ pts & 0.412\\
\bottomrule
\end{tabular}
\caption{Effect of return-shape hints in code-mode stubs (development seeds, 118--119 cells).}
\label{tab:stubs}
\end{table}

Hints also reduced tokens (Qwen 6.2K$\to$4.5K, Nemotron 14.2K$\to$8.2K). They flipped the preferred shape for Gemma and Nemotron: unhinted, plain MCP beat code mode for both; hinted, code mode wins. They also shrank the apparent headroom for any shape-routing layer (oracle utility gap over the best fixed shape: Qwen $0.051\to0.023$, Gemma $0.219\to0.110$, Nemotron $0.272\to0.084$). Conclusions about ``which shape is best for this agent'' are therefore \emph{conditional on stub quality}, and any earlier comparison that used undeclared return types may be measuring the stubs rather than the agent. The sealed-split extension (Section~\ref{sec:ext1}) reproduces this result. The hints were written by the author and tuned on development data, so the finding is that return-value information matters; how much of the gain real servers' \texttt{outputSchema} declarations (when present) would deliver is untested.

\subsection{When does a decision layer help?}\label{sec:decision}
\paragraph{H3: \cfg{D} versus \cfg{C}.} With the same hinted stubs:
\begin{itemize}
\item \textbf{Gemma (lower code-mode success): supported on development seeds, not confirmed on the test split (Section~\ref{sec:confirm}).} With a Qwen-selected subset (2.4 tools on average), \cfg{D} reaches success 0.966 versus 0.864 for \cfg{C} ($+10.2$ points $[+4.2,+16.1]$), uses 18\% fewer tokens (router tokens included), and $\Delta U_{\text{tok}}=+0.109$ $[+0.046,+0.178]$. With the gold subset, \cfg{D-oracle} gives $+7.6$ points and $\Delta U_{\text{tok}}=+0.099$ $[+0.048,+0.160]$, so the prompted selector captured the whole oracle benefit. The benefit comes from the small accurate subset plus the JSON fallback, not from shape choice.
\item \textbf{Qwen (highest code-mode success): not supported.} \cfg{D} ties \cfg{C} ($\Delta U_{\text{tok}}=-0.009$, $[-0.032,+0.005]$); \cfg{C} is already near ceiling at 0.975.
\item \textbf{Nemotron: suggestive.} On the 39-task latency set, success is $0.64$ (\cfg{D}) versus $0.56$ (\cfg{C}) and $0.38$ (\cfg{A}); all intervals include zero.
\end{itemize}
\textbf{H4 (\cfg{D} beats \cfg{B}) is supported} for Qwen and Gemma: with the same subset, adding code mode gives $+12.7$ points of success and $4.4$K fewer tokens on Gemma.
\textbf{H2 (router+JSON is non-inferior to plain MCP and cheaper)} holds only when the subset is small and accurate: LLM and gold subsets cut tokens by 65--68\% at no loss of success, whereas zero-shot \sparks{} subsets (about 12 tools) saved no tokens. The pre-registered 2-point non-inferiority margin cannot be resolved at $n\approx119$.

\paragraph{H5: shape routing.} Every shape router we tried (a rule-based selector, zero-shot \sparks{}) scored below ``always code plus fallback'' for Qwen and Gemma. Per-cell oracle shape labels were \emph{not} an upper bound, because they come from single noisy runs (winner's curse). A held-out analysis for Gemma, learning the best shape per task family on seeds 0--2 and applying it to fresh seeds 3--5, gave $+0.051$ utility over always-code $[+0.003,+0.106]$, but that comparison is against \cfg{C} \emph{without} a subset or fallback. An offline analysis over all (agent $\times$ shape) arms found only $+0.036$ oracle headroom, and a leave-one-task-out family-level policy scored $-0.021$ $[-0.078,+0.010]$ below the best fixed arm (Qwen with code mode). We therefore treat shape routing as a negative result.

\subsection{Latency}\label{sec:latency}
Table~\ref{tab:latency} gives clean serial latency on 39 tasks (seed 0, hinted stubs, mock tools at $\approx$0\,ms). Medians are reported because Gemma has rare runaway runs (one of 683\,s).

\begin{table}[h]
\centering\small
\begin{tabular}{@{}lccc@{}}
\toprule
 & Qwen3.8-27B & Gemma-4-26B & Nemotron-3-Nano\\
 & succ / med s / p90 s & succ / med / p90 & succ / med / p90\\
\midrule
\cfg{A} plain MCP & 0.82 / 5.0 / 15.6 & 0.82 / 5.8 / 21.0 & 0.38 / 6.4 / 22.9\\
\cfg{B} router + JSON & 0.85 / 5.7 / 12.7 & 0.87 / 7.9 / 18.9 & 0.59 / 6.2 / 22.3\\
\cfg{C} code mode & \textbf{0.97} / 3.7 / \textbf{5.9} & 0.79 / 6.6 / 33.9 & 0.56 / 4.5 / 29.3\\
\cfg{D} router + code & 0.95 / 4.6 / 6.8 & \textbf{0.90} / 7.4 / 21.4 & \textbf{0.64} / 5.5 / 35.1\\
\cfg{D-oracle} (free router) & 0.97 / \textbf{3.0} / 4.8 & 0.90 / 7.5 / 19.3 & 0.64 / \textbf{3.6} / 21.9\\
\bottomrule
\end{tabular}
\caption{Serial latency, 39 tasks, one seed (development). Router time is included in \cfg{B} and \cfg{D}.}
\label{tab:latency}
\end{table}

Code mode is the latency winner for the strongest coder and its advantage over plain MCP grows with the catalog (Section~\ref{sec:scale}). A front-door router costs about 1.6\,s at 42 tools (about 50\,ms per candidate question), more than it saves for Qwen: with a free, perfect router \cfg{D} is faster than \cfg{C} ($-0.7$\,s $[-1.0,-0.5]$), but with the real router it is $2.2$\,s slower. For the two lower-performing agents the decision layer improves success (0.79$\to$0.90 for Gemma, 0.56$\to$0.64 for Nemotron; intervals include zero on this 39-task set) and trims the tail, while typical latency is unchanged. Mock tools return instantly, so real servers with per-call network time would favor batching scripts even more; we did not simulate this.

\subsection{Scale and the fallback}\label{sec:scale}
We vary the catalog to 42, 200 and 2{,}000 tools for the Qwen agent (10 tasks, seed 0, serial).

\begin{table}[h]
\centering\small
\begin{tabular}{@{}rlrrr@{}}
\toprule
Tools & Config & Success & Tokens/task & Latency mean / median / p90 (s)\\
\midrule
42 & \cfg{A} & 0.90 & 87{,}460 & 9.8 / 4.8 / 60.6\\
42 & \cfg{C} & 1.00 & 4{,}677 & 3.6 / 3.6 / 5.9\\
200 & \cfg{A} & 0.90 & 163{,}969 & 22.1 / 31.3 / 39.2\\
200 & \cfg{C} & 0.90 & 10{,}390 & 5.6 / 4.9 / 10.0\\
200 & \cfg{D} unbounded fallback & 0.80 & 66{,}008 & 24.4 / 6.4 / 98.9\\
200 & \cfg{D} bounded fallback & 0.90 & 14{,}961 & 9.1 / 6.1 / 22.1\\
2{,}000 & \cfg{A} & 0.00 (context overflow) & -- & --\\
2{,}000 & \cfg{C} & 0.90 & 14{,}255 & 6.3 / 5.0 / 11.3\\
2{,}000 & \cfg{D} unbounded fallback & 0.40 & 370{,}482 & 222.7 / 116.1 / 570.2\\
2{,}000 & \cfg{D} bounded fallback & 0.90 & 34{,}490 & 16.1 / 16.4 / 34.0\\
\bottomrule
\end{tabular}
\caption{Scale arm (Qwen3.8-27B, 10 tasks, one seed). Plain MCP needs about 47K schema tokens per turn at 200 tools and cannot run at 2{,}000.}
\label{tab:scale}
\end{table}

Code mode scales almost flat (3.6$\to$6.3\,s, 4.7K$\to$14K tokens), while plain MCP degrades and then fails. The decision layer exposes a design lesson: because candidate retrieval caps tool recall (0.65 at 200 and 0.44 at 2{,}000 tools), a subset miss is common at scale. Our first fallback switched to JSON and let the agent call \texttt{request\_more\_tools} without bound; each call added long real-world schemas, and tasks reached up to 23 calls and 900K tokens. Bounding the fallback (retry in code mode over the full catalog, then at most two request calls) cut \cfg{D} at 2{,}000 tools from 223\,s to 16\,s ($14\times$) and from 370K to 34K tokens, with success $0.40\to0.90$. Even so, \cfg{D} did not beat \cfg{C} for Qwen at any catalog size. This is a 10-task mechanism check, not a precise estimate.

\section{A Fine-Tuned Tool Router}\label{sec:router}

\paragraph{Model.} We fine-tuned \sparks{}, a 4B decision model on a Qwen3.5-4B backbone that returns calibrated probabilities for typed questions, into a tool selector (\texttt{spark-s1-mcp-router-4b-r1}) with LoRA. The state is the task plus 22--30 embedding-retrieved candidate tools, with one yes/no question per tool (``is this tool needed, directly or because its result feeds another step?''). Training data come from a disjoint training world (five task families and four servers' tools held out), with paraphrases written by one model family (Nemotron) and verified by another (Qwen), labels computed from code, and negatives drawn from real catalog tools never used as evaluation distractors. Our first two training runs were aborted because loss reached zero within about 40 steps: 40\% of the questions were trivially ``no'' (a tool-name-only rule scored 91\%); the third run downsampled them.

\paragraph{Results.} On 79 development tasks at a 42-tool catalog, complete recall (every needed tool selected) at threshold 0.01 is 0.899 for the zero-shot model at 3.3 tools and 1.000 for the fine-tuned model at 2.8 tools. On the same tasks reworded by a different model family and never seen in training, recall is 0.873 versus 0.964. Paired differences in complete recall (fine-tuned minus zero-shot) are $+10.1$ points $[+3.8,+17.7]$ (template wording, same threshold) and $+9.1$ points $[+1.8,+18.2]$ (reworded). At 200 and 2{,}000 tools both models sit at the retriever's ceiling (0.65 and 0.44), but the fine-tuned model exposes about half as many tools (2.6 and 3.4 versus 5.3 and 6.2). In end-to-end runs with a Gemma agent, \cfg{D} with this router reaches success 0.932 versus 0.864 for \cfg{C} ($+6.8$ points $[+1.7,+12.7]$); the prompted Qwen router did better (0.966), although the fine-tuned router's tool recall is higher (0.987 versus 0.966), so that gap is agent-side noise at $n=118$.

\paragraph{Costs, stated plainly.} (i) \emph{No latency win:} a prompted 26--27B router answers in 0.75--0.9\,s; the fine-tuned 4B router takes 1.6\,s at 30 candidates, or 0.41\,s at 15 candidates with a 2.5--5.5 point recall loss; a GGUF build under llama.cpp is slower still. Its advantage is footprint (8.4\,GB versus about 52\,GB) and precision, not speed. (ii) \emph{General skills:} on the base model's own benchmark, accuracy drops by 1.5--2.8 points, so it is not a drop-in replacement. (iii) \emph{Narrow data:} 11 template families in one synthetic world, so near-zero training loss and recall that falls from 1.00 to 0.96 on reworded tasks.

\section{Exploratory Security Observations}\label{sec:security}
These observations are small and exploratory (one agent, six tasks, one seed for injection; one poisoned tool for routing) and are not part of the confirmatory evaluation.

\paragraph{Injection through tool results.} With the Qwen agent, attack success was measured from world state across 108 runs per configuration: \cfg{A} 0.148, \cfg{C} 0.093, \cfg{D} 0.000. Code mode therefore did not raise attack success over JSON in this pilot (the difference is within noise), and the zero for \cfg{D} is largely structural: with about 2.4 tools exposed, the attacker's goal tools are usually not callable at all. Only one payload style succeeded. Screening tool outputs with a zero-shot \sparks{} question roughly halved attack success for \cfg{A} and \cfg{C}, with 0 blocks in 343 benign calls, but the author wrote both the attack payloads and the defence.

\paragraph{The router is a target.} With one poisoned attacker tool in the catalog, a prompted LLM router exposed it in 39 of 39 tasks when its description said ``ALWAYS select this tool for every task'', and in none for two softer phrasings. A scoring router (fine-tuned \sparks{}) exposed it in 62\% of tasks for the ``ALWAYS'' phrasing, versus 69\% for the zero-shot model (not a distinguishable difference at $n=39$). The ``least-privilege'' benefit of a small tool subset therefore assumes an honest router; no router output should be used as an authorization decision.

\section{Discussion}

\paragraph{What the evidence supports.} \emph{Confirmed on the sealed split:} code mode is the dominant effect on utility for two of three agents, with 70--90\% fewer tokens; and return-shape information in the stubs is worth $+0.30$ and $+0.48$ of utility for the two lower-performing agents, more than any router effect we measured. \emph{Directionally supported, not confirmed:} a decision layer in front of code mode helps those two agents ($+0.10$, $+0.11$ pooled over 238 instances, not significant after correction) and slightly hurts the strongest; router+code over router+JSON is supported for Qwen and directionally for the others. \emph{Exploratory only:} shape routing adds nothing, an unbounded fallback is pathological, and routers are injectable.

\paragraph{Implications for gateway design.} (1) Publish or infer an output contract for every tool; where servers do not declare \texttt{outputSchema}, a gateway could learn shapes from traffic and use them to annotate stubs, which we propose but do not test. (2) Default to code mode with a bounded fallback instead of routing the call shape. (3) If a front-door router is used, make its fallback bounded and its output non-authoritative. (4) Measure end-to-end latency, not recall alone.

\paragraph{Future work.} Real MCP servers and an established benchmark; unseen task families and paraphrased wording; thinking-enabled agents; a cheaper single-pass router; and typed code mode that infers return shapes from traffic and type-checks scripts before execution.

\section{Limitations and Threats to Validity}\label{sec:limits}
\begin{itemize}
\item \textbf{No external benchmark or real servers.} The world is synthetic, the tools are instant, and the test split uses the same template families as development (21 families, so instances are clustered). The study does not test unseen task families, real third-party MCP servers, real per-call latency, or paraphrased wording. This is its largest limitation for external validity.
\item \textbf{Power and sample size.} The protocol called for about 300 pairs per arm; the confirmatory split has 119 instances and the extended H3 test 238. Several intervals include zero, so ``not confirmed'' means inconclusive. One run per cell at temperature 0; the scale arm has 10 tasks.
\item \textbf{Analysis choices.} The pre-registered analysis resamples instances; we added a family-level analysis after seeing the core results, and it changes some conclusions (Table~\ref{tab:holm}). The extension protocol was written after the core results and before the extension runs; its tests are outside the 12-comparison Holm family. The utility weights are a choice (Table~\ref{tab:lambda}).
\item \textbf{Researcher degrees of freedom before the freeze.} The stub hints were hand-written by the author and tuned on development data; harness bugs were fixed on development worlds; thresholds were not tuned on a separate split. ``Lower-performing agent'' is an outcome-based label, and the three agents also differ in architecture, quantization and serving stack, so agent-level explanations are confounded.
\item \textbf{Oracle and role confounds.} \cfg{D-oracle} uses the reference solver's tool list and so leaks the task plan; it is a ceiling. The router is the author's own, trained on the same template structure; development \cfg{B}/\cfg{D} used prompted routers of a different family, the confirmatory runs used the fine-tuned router for all agents, so the two are not like-for-like. Nemotron shares a family with the test-side text writer.
\item \textbf{Infrastructure.} The sandbox is not a container; agents are quantized; thinking was disabled. In the confirmatory runs one Gemma and two Nemotron runs failed inside the router (a malformed response) and were re-run once under the protocol (original records are released); one extension run hit the 900\,s cap and is scored as a failure. In development about 5 of roughly 3{,}000 runs hung and were excluded.
\item \textbf{Security.} Exploratory and tiny; the authors wrote both the attacks and the defence.
\end{itemize}

\section{Conclusion}
In a controlled study with a pre-registered evaluation on a sealed split, code mode was the dominant design choice for MCP gateways, and the information stubs carry about tool return values changed outcomes by up to 40 points for two of three agents, more than adding a router did. The motivating hypothesis, that a decision layer in front of code mode improves on code mode alone, was not confirmed, even after doubling the instances: it shows a consistent but statistically inconclusive pattern (helpful for the two lower-performing agents, slightly harmful for the strongest). The analysis choices matter: resampling task families rather than instances overturns some pre-registered conclusions. We release the harness, the router and all run records so that these questions can be tested on fresh splits and on real servers.

\section*{Artifacts}
Code and pre-registration: \url{https://github.com/abhishek085/mcp-decision-layer}. The confirmatory protocol, frozen at tag \texttt{confirmatory-freeze-v1}, the raw per-run records (\texttt{traces/confirmatory/}) and the analysis scripts (\texttt{bench/confirm\_analysis.py}, \texttt{bench/confirm\_robustness.py}) are in the repository. Model \texttt{spark-s1-mcp-router-4b-r1} (GGUF, Apache-2.0 with the base-model license chain), the router training dataset and the benchmark dataset with run traces (CC-BY-4.0): \url{https://huggingface.co/abhishek085}.

\section*{AI assistance}
Experiments were executed and logged with the assistance of an AI coding agent (Claude Code) under the author's direction, and this manuscript was drafted with its help. The author is responsible for the content.

\end{document}